\documentclass[aps,prb,twocolumn,superscriptaddress,showpacs,longbibliography]{revtex4-1}
\usepackage{color}
\usepackage[normalem]{ulem}
\usepackage{soul}
\usepackage{amsmath}
\usepackage{graphicx}
\usepackage[colorlinks]{hyperref}
\usepackage{cancel}
\hypersetup{
    unicode=false,          
    pdftoolbar=true,        
    pdfmenubar=true,        
    pdffitwindow=false,     
    pdfstartview={FitH},    
    pdftitle={Measuring the imaginary time dynamics of quantum materials},    
    pdfauthor={S. Lederer},     
    pdfsubject={Subject},   
    pdfcreator={Creator},   
    pdfproducer={Producer}, 
    pdfkeywords={keyword1} {key2} {key3}, 
    pdfnewwindow=true,      
    colorlinks=true,       
    linkcolor=blue,          
    citecolor=blue,        
    filecolor=blue,      
    urlcolor=blue,       
    plainpages=false        
}

\def\be{\begin{equation}}       \def\ee{\end{equation}}
\def\bea{\begin{eqnarray}}      \def\eea{\end{eqnarray}}
\def\ba{\begin{array} }
\def\ea{\end{array} }
\def\bnum{\begin{enumerate} }
\def\enum{\end{enumerate}}

\def\nn{\nonumber}

\def\=>{\Rightarrow}
\def\>{\rightarrow}

\def\eye2{Fathbb{I}}

\def\d0{\Delta_{0}}

\def\Co122{${\rm Ba(Fe_{1-x}Co_x)_2As_2}$}

\newcommand{\eq}[1]{\begin{align}#1\end{align}}

\begin{document}
\title{Measuring the imaginary time  dynamics of  quantum materials}
\author{S. Lederer}
\affiliation{Cornell University, Ithaca, New York 14850, USA}
\author{D. Jost}
\affiliation{Walther Meissner Institut, Bayerische Akademie der Wissenschaften,
85748 Garching, Germany}
\affiliation{Fakult\"at f\"ur Physik E23, Technische Universit\"at M\"unchen,
85748 Garching, Germany}
\author{T. B\"ohm}
\altaffiliation{Present address: TNG Technology Consulting GmbH, Beta-Strasse, 85774 Unterf\"ohring, Germany}
\affiliation{Walther Meissner Institut, Bayerische Akademie der Wissenschaften,
85748 Garching, Germany}
\affiliation{Fakult\"at f\"ur Physik E23, Technische Universit\"at M\"unchen,
85748 Garching, Germany}
\author{R. Hackl}
\affiliation{Walther Meissner Institut, Bayerische Akademie der Wissenschaften,
85748 Garching, Germany}
\author{E. Berg}
\affiliation{Department of Condensed Matter Physics, The Weizmann Institute of Science, Rehovot, 76100, Israel}
\author{S. A. Kivelson}
\affiliation{Department of Physics, Stanford University, Stanford, CA 94305, USA}

\date{\today }

\begin{abstract}
Theoretical analysis typically involves imaginary-time correlation functions.
Inferring real-time  dynamical response functions from this information is notoriously difficult.  However, as we articulate here, it is straightforward to compute imaginary-time correlators from the {\em measured} frequency dependence of (real-time) response functions.  In addition to facilitating comparison between theory and experiment, the proposed approach can be useful in extracting certain aspects of the (long-time relaxational) dynamics from a complex data set. We illustrate this with an analysis of the nematic response inferred from Raman scattering spectroscopy on the iron-based superconductor \Co122, { which includes a new method for identifying a putative quantum critical contribution to that response.}
\end{abstract}

\maketitle
Spectroscopic probes provide a wealth of information about the dynamics of quantum systems. Spectra are often very complicated.  
Sometimes, however, individual spectral features are of less interest than the overall evolution of the spectrum as parameters (such as temperature, pressure, doping, or magnetic field), are varied. In this setting, it is necessary to condense the considerable information in each spectrum into a few numbers. We propose a new method 
for doing this based on the computation of correlation functions in the \emph{imaginary} time domain. The method is unbiased, numerically reliable, and allows unambiguous comparison with the results 
of state of the art numerical methods.

It goes without saying that laboratory experiments measure real-time correlators.  However, many theoretical methods that have been deployed to extract non-perturbative results on strongly interacting quantum systems, including various forms of quantum Monte Carlo studies, work exclusively in imaginary time.  Given the uncertainties in analytic continuation, one reason to compute imaginary time correlators from laboratory data is that it transforms it into a form that can be directly compared with this class of theoretical results.

The transformation from real frequency to imaginary time, discussed in further detail below, discards much of the rich information present in real-time data. For instance, a well defined normal mode (quasiparticle) with an energy $\epsilon \gg k_BT$ (with $T$ the temperature) shows up as a sharp peak in an appropriate real frequency response function, but the corresponding feature in the imaginary-time correlator varies as $\exp[-\epsilon\tau/\hbar]$, and so makes no contribution to long-time properties. On the other hand, the long time relaxational dynamics of a system -- the dynamics that control its approach to equilibrium -- typically dominate the long-imaginary time dynamics as well.  Thus, using measured response functions to compute  long-imaginary-time behavior  of the corresponding correlators can be viewed as a method of intrinsic and unbiased filtering, which extracts certain interesting information from a complex spectral response.

In principle, it is possible to compute the real-time (or frequency, $\omega$) fluctuational dynamics of any system in equilibrium from  imaginary time (or Matsubara frequency, $\omega_n$) correlation functions and vice versa.  In practice, inferring real time dynamics from imaginary time data involves an analytic continuation that can rarely be carried out without additional assumptions.  This ambiguity follows from the fact that the discrete Matsubara frequencies, $\omega_n$, have spacing $\Delta\omega\equiv \omega_{n+1}-\omega_n = 2\pi k_BT/\hbar$, making features which vary as a function of $\omega$ more rapidly than $\Delta\omega$  difficult to discern in the imaginary time response functions.  
It is, however, straightforward to compute imaginary time correlation functions from measured real frequency quantities.

Here, we give explicit formulas for computing imaginary time correlators from response functions measurable in the laboratory.  Building on the work in Ref. \onlinecite{Randeria1992,Trivedi1995,Trivedi1996}, we treat explicitly the general case of linear response of (bosonic) physical observables,  as well as the electron spectral function (measurable in tunneling and photo-emission spectroscopy).  To illustrate what information is emphasized and what is suppressed, we carry out this program for various simple and physically plausible  assumed forms of a response function.  Finally, 
 to illustrate the usefulness of the approach, we take high resolution Raman data measured on Ba(Fe$_{1-x}$Co$_x$)$_2$As$_2$ in the $B_{2g}$ channel\cite{footnote} and compute the corresponding imaginary time correlator, which yields a sharp diagnostic of the structural transition in that channel.  
 
\section{Computing imaginary-time correlators from  spectroscopic measurements}
\subsection{Dissipative linear response functions}
For any observables $\Phi_a$ and $\Phi_b$, it follows from linear-response theory and the fluctuation dissipation theorem that there is a relation between the dissipative part of the linear response function, $\chi_{ab}^{\prime\prime}(\omega)$, and the imaginary-time-ordered correlation function, $\tilde\Lambda_{ab}(\tau)$:
\be
\tilde\Lambda_{ab}(\tau) = \int \frac {d\omega}{2\pi}\ \chi^{\prime\prime}_{ab}(\omega)\ \left[\frac {  \exp(\omega[\tau-\beta/2])}{\sinh(\beta\omega/2)}\right],
\label{basic}
\ee
where (in units in which $k_B=\hbar=1$) $ \chi^{\prime\prime}$ is the Fourier transform of
\be
\tilde \chi_{ab}^{\prime\prime}(t) \equiv \frac 1 2 \langle [\Phi_a(t),\Phi_b(0)]\rangle,
\ee
and, for $0 \leq \tau\leq \beta$
\be
\tilde\Lambda_{ab}(\tau)\equiv \langle \Phi_a(-i\tau)\Phi_b(0)\rangle,
\ee
with $\beta=1/T$. 
The relation between $\chi^{\prime\prime}$ and the imaginary time correlation function in the Matsubara frequency domain is
 \be
\Lambda_{ab}(\omega_n) = \int \frac {d\omega}{\pi} \ \chi^{\prime\prime}_{ab}(\omega)\ \left[\frac {\omega}{\omega^2 + \omega_n^2}\right],
\label{matsubara}
\ee
where $\omega_n=2\pi n T$. {Because $\tilde \Lambda_{ab}(\tau)$ is a bosonic correlator, $\tilde \Lambda_{ab}(\tau) = \tilde
\Lambda_{ba}(\beta-\tau)$ .  Thus, if we are interested in the ``long-time'' behavior of $\tilde \Lambda_{ab}$, we mean we are
interested in the longest-possible times, {\it i.e.} $\tau\approx \beta/2$.}  The important point to note about Eq. \ref{basic} is
that for $\tau \approx \beta/2$, the integral is dominated by the range of frequencies $|\omega| \lesssim T$, so {\em the long
imaginary-time dynamics can be computed from  measurements of the response function in a very limited range of frequencies.}

As one important example, let $\Phi_a$ be a component of the electrical current operator, whose associated susceptibility is
proportional to the conductivity. Let $\sigma^\prime_{aa}(\omega)$ be the real part of the optical conductivity, and $\tilde
\Lambda_{aa}(\tau)$ be the imaginary time ordered current-current correlator.  Here $a$ is a tensor index indicating a spatial
direction. The Kubo formula relates the conductivity to $\chi^{\prime\prime}$, and consequently\cite{Trivedi1996}
\be
\tilde \Lambda_{aa}( \tau) = \int \frac {d\omega}{2\pi}\ \omega\sigma^\prime_{aa}(\omega)\ \left[\frac {  \cosh[\omega(\beta/2-\tau)]}{\sinh(\beta\omega/2)}\right].
\label{sigma}
\ee

The other case we treat here is where $\Phi_a$ 
is an order parameter field. For instance, $\Phi_a$ could be a component of the spin density at an appropriate ordering vector
$\vec Q$,  so that the resulting susceptibility (which has a singular response near a magnetic transition) can be measured in
inelastic neutron scattering. If $\Phi_a$ is a component of the fermion quadrupole density in some symmetry channel ($B_{1g}$,
$B_{2g}$, etc.), then the resulting susceptibility (which has a singular response near a nematic transition), can be measured in non-resonant Raman scattering\cite{shastry1990}.

\subsection{Electronic spectral function}
Similar expressions 
relate the imaginary-time-ordered Green function, $\tilde G(\vec k,\tau)$ to the single particle spectral function, $A(\vec
k,\omega)\equiv -1/\pi Im({\cal G}(\vec k,\omega))$, where $\cal G$ is the real frequency (retarded) Green function\cite{Trivedi1995}.
\bea
\label{A}
\tilde G(\vec k,\tau) =&& \int  {d\omega}\ A(\vec k,\omega)\ \left[\frac {  \exp[\omega(\beta/2-\tau)]}{2\cosh(\beta\omega/2)}\right], \\
=&& \int  {d\omega}\ I(\vec k,\omega)\ \exp[\omega(\beta- \tau)]
\nonumber
\eea
where $\vec k$ is the momentum (or Bloch wave-vector), we have assumed $\tau$ in the range $0\leq \tau < \beta$, and $I(\vec
k,\omega) \equiv f(\omega) A(\vec k,\omega)$ is the occupation-weighted spectral function (as measured in ARPES), where
$f(\omega)= [e^{\beta \omega}+1]^{-1}$ is the Fermi function.  Again, except at very short imaginary times, the imaginary time correlator can be readily computed from 
the experimentally measured response function over a range of frequencies of order $T$ about the Fermi energy.

In Appendix \ref{sec:examples} we explicitly derive the transformations for a few special cases relevant for correlated
systems including the marginal Fermi liquid\cite{varma1989phenomenology} and power-law scaling close to a quantum critical point.

\section{
The imaginary time quadrapolar correlations in  \Co122}
Here we apply the proposed analysis to the  
experimentally measured temperature ($T$) and doping ($x$) dependent
 Raman response of  \Co122, with particular focus on the critical electronic quadrapolar fluctuations in the vicinity of the structural (nematic) transition at $T_s$.

\subsection{The static susceptibility and $\tilde\Lambda(\beta/2)$}
Raman scattering measures a dissipative response, and can therefore yield the imaginary part $\chi''(\omega,T)$ of an appropriate
susceptibility. The static value of the real part of the susceptibility, $\chi'(0,T)$, is related to $\chi''(\omega,T)$ by the Kramers-Kronig transformation,
\bea
\chi'(0,T)=\int \frac{d\omega}{\pi} \frac{\chi''(\omega,T)}{\omega}.
\label{eq:kk}
\eea
Note that Eq. (\ref{matsubara}) reduces to Eq. (\ref{eq:kk}) when $\omega_n=0$, so that $\Lambda(0)=\chi'(0,T)$. In practice,
$\chi'(0,T)$ can almost never be precisely determined from Raman measurements, because the integrand falls off too weakly at large
frequency, necessitating an arbitrary cut-off procedure;  { this is a generic problem with Kramers-Kronig analysis.}  As an alternative, one can use the same data to determine the value of the imaginary time correlator at time $\beta/2$, via
\bea
\tilde\Lambda(\beta/2)=\int \frac{d\omega}{2\pi}\frac{\chi^{\prime\prime}(\omega)}{\sinh(\beta\omega/2)}.
\label{eq:betaover2}
\eea
Since $|\sinh(x)|\geq |x|$ for all x (with the inequality saturated as $x\to 0$), we have the following inequality,
\bea
\tilde\Lambda(\beta/2) \leq T \int \frac{d\omega}{\pi}\frac{\chi^{\prime\prime}(\omega)}{\omega}=T\chi^\prime(0).
\label{eq:chi'(0)}
\eea
Evidently $\tilde \Lambda(\beta/2)/T$ is bounded above by the static susceptibility, with the bound nearly saturated when spectral weight
is concentrated at frequencies $\omega\ll T$. In fact, $\tilde \Lambda(\beta/2)/T$ contains the same universal information as the
static susceptibility under a wide range of assumptions. For instance, at a continuous phase transition at nonzero temperature,
$\tilde \Lambda(\beta/2)/T$ has the same divergent behavior as the static susceptibility. This can be seen by writing the quantity in Fourier transform:
\bea
\tilde \Lambda(\beta/2)=&T\sum_{n} e^{-i\nu_n\beta/2}\Lambda(\nu_n)\nn\\
=&T\sum_{n} (-1)^n \Lambda(\nu_n)\nn\\
=&T\chi'(0,T)+\dots,
\label{eq:bound}
\eea
where we have used the fact that $\Lambda(0)=\chi^\prime(0,T)$, and dots refer to the contribution from nonzero Matsubara frequencies
which, per Eq. \ref{matsubara}, are insensitive to the asymptotically low frequencies at which critical behavior in
$\chi^{\prime\prime}$ is found. At a quantum critical point obeying $\omega/T$ scaling, $\tilde\Lambda(\beta/2)/T$ also has the same
divergence as the static susceptibility in the low temperature limit.

A key practical advantage of $\tilde\Lambda(\beta/2)/T$ as a measure of low frequency fluctuations is the fact that the $\sinh(\beta\omega/2)$ in the denominator of  Eq. \ref{eq:bound} 
yields an {exponential cutoff} at high energies the scale of which is given by temperature. This means that
$\tilde\Lambda(\beta/2)/T$, unlike $\chi^\prime(0)$, is subject to essentially no error due to a lack of knowledge of high frequencies.
As we will show, it is therefore a valuable and unambiguous method of analysis for Raman spectra.

\subsection{Raman spectra of Co-doped BaFe$_2$As$_2$}
We demonstrate now the effect of using the dimensionless imaginary-time correlation function $\beta{\tilde\Lambda(\beta/2,T)}$ for the analysis of Raman spectra and put it into perspective with other methods for extracting properties 
{in the low frequency limit}. In particular, we compare the results obtained for $\beta{\tilde\Lambda(\beta/2,T)}$ with the static Raman susceptibility $\chi^\prime(0,T)$.

To this end we 
{ have extended earlier measurements\cite{Gallais2013,Kretzschmar2016}  of the Raman spectra of \Co122 both to obtain data over a wider range of energies, and to access a finer grid of temperatures.}
By extending the range of frequencies up to 1000\,cm$^{-1}$  
we ensure that the range is sufficient to unambiguously determine $\beta{\tilde\Lambda(\beta/2,T)}$  even at room temperature, where 1000\,cm$^{-1}$ is $4.7\,k_B T$. 
 The 
 dense average grid of temperatures, $\bar{\Delta T} = 23$\,K, is  needed to identify critical behavior in the neighborhood of the structural transition temperature, $T_s$.

\begin{figure}[ht]
   \centering
     \includegraphics[width=8.5cm]{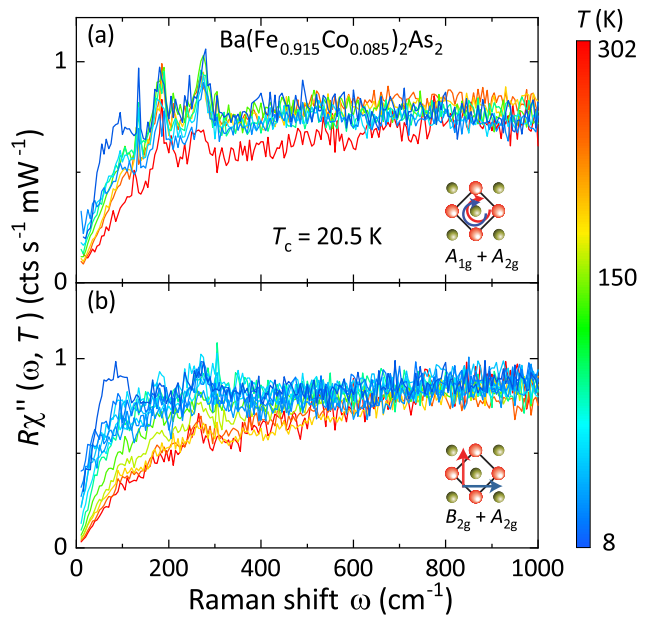}
   \caption{{Raw data of the Raman response $R\chi^{\prime\prime}(\Omega,T)$ of ${\rm
   Ba(Fe_{0.915}Co_{0.085})_2As_2}$. We use a continuous color scale for the temperature (right scale bar). The (a) $A_{1g}+A_{2g}$ and
   (b) $B_{2g}+A_{2g}$ spectra are measured in $RR$ ($R=(x+iy)/\sqrt{2}$) and $xy$ polarization, respectively, where
   $x$ and $y$ are the axes of the 2\,Fe crystallographic cell in the tetragonal phase as indicated pictorially. 
     }
   }
   \label{fig:BFCA85}
 \end{figure}

\begin{figure}[ht]
   \centering
     \includegraphics[width=8.5cm]{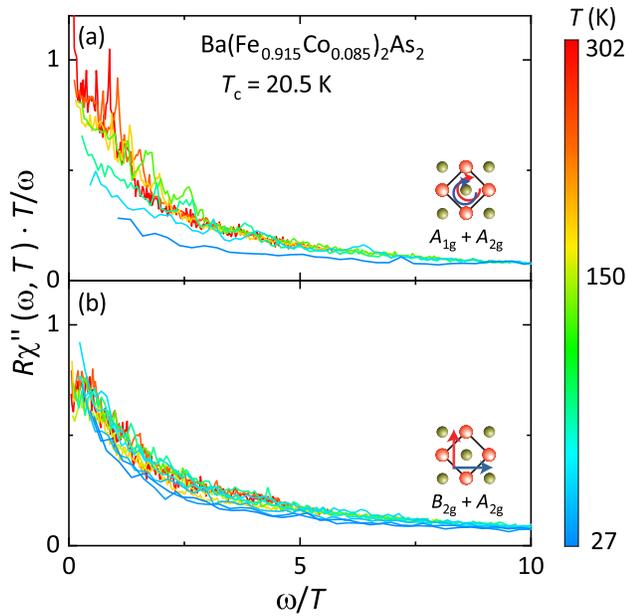}
   \caption{{Same data as in Fig. \ref{fig:BFCA85}, but now in terms of scaled variables.  The $y$ axis is $\chi^{\prime\prime}(\omega,T) \cdot T/\omega$ and the $x$ axis is $\omega/T$.  Only data for $T> T_c=$20.5K is shown.
     }
   }
   \label{scaled}
 \end{figure}

Fig.~\ref{fig:BFCA85} shows the Raman spectra of overdoped ${\rm Ba(Fe_{0.915}Co_{0.085})_2As_2}$ 
in the
 $A_{1g}+A_{2g}$  channel [essentially s-wave, panel (a)] and the $B_{2g} + A_{2g}$ [{ d-wave-like}, panel (b)].
 The spectra are constant and temperature independent (to within $\pm 5\%$) at energies above 700\,cm$^{-1}$. Below
 700\,cm$^{-1}$ the intensity increases  upon cooling,  with the $B_{2g}+A_{2g}$ spectra (Fig.~\ref{fig:BFCA85}b) displaying a slightly stronger variation. 
{ Fig. \ref{scaled} shows the same data, but now presented as a function of scaled variables.  Sufficiently close to certain quantum critical points, one  expects  critical response functions to exihibit $\omega/T$ scaling, which would mean that scaling the data at various $T$ and $\omega$ as in the figure would collapse the data onto a single curve.  The  data in the B$_{2g}$ channel shows an approximate version of such a scaling collapse;  the A$_{1g}$ data somewhat less so.}

\begin{figure}
  \centering
  \includegraphics [width=8.5cm]{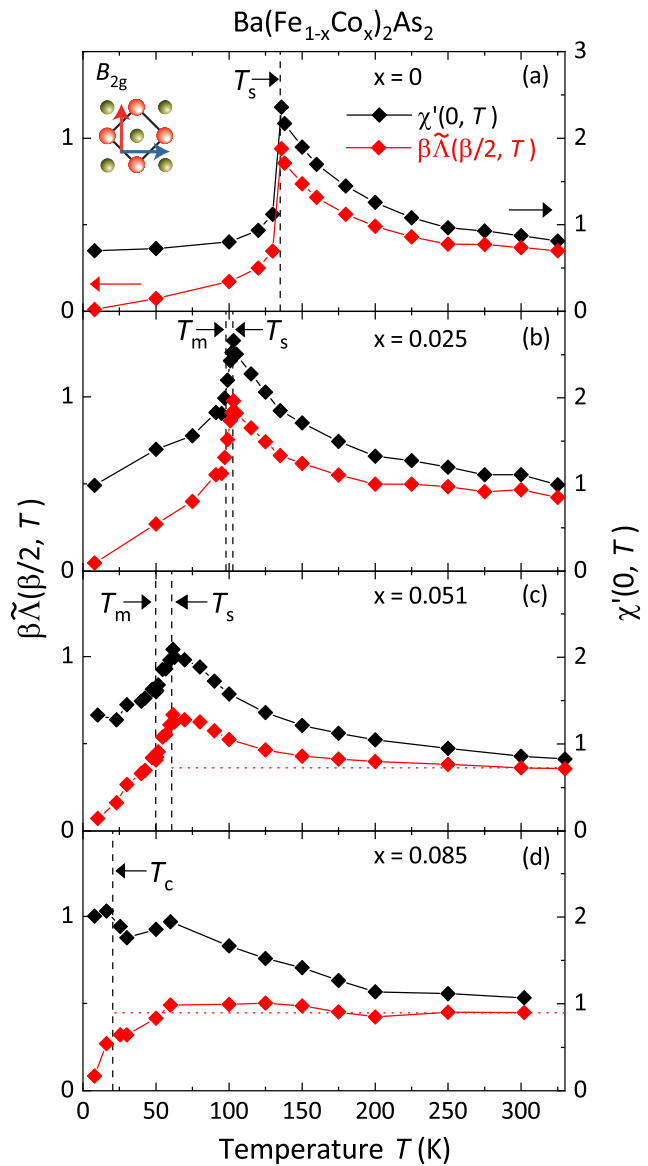}
  \caption{Temperature dependence of two different measures of the low frequency Raman response of \Co122 in the $B_{2g}$ symmetry
  channel. The doping concentration $x$ determines the distance from a putative QCP at $x_c\approx 0.06$ between 0.051 and 0.085. $T_s$, $T_m$ and $T_c$ are the structural, magnetic and superconducting transition temperatures, respectively. At $x=0$  $T_s$ and $T_m$ coincide. Shown in black (right axis) is the static susceptibility $\chi'(0, T)$, computed from the measured
  $\chi^{\prime\prime}(\omega,T)$ using the Kramers-Kronig relation of Eq.~\ref{eq:kk}, integrated up to a cutoff frequency of
  1000\,cm$^{-1}$ (approximately 124\,meV). In red (left axis) is the suitably scaled imaginary time correlation function $\tilde
  \Lambda(\beta/2,T)$, defined in Eq.~\ref{eq:betaover2}. Both quantities capture the singular temperature dependence of the
  Raman response near the structural transition, but the imaginary time correlator requires no manual cutoff procedure. The two
  quantities differ most substantially near zero temperature, where $\tilde\Lambda(\beta/2,T)$ must vanish, since it only captures the
  dynamics at frequencies of order the temperature. The Raman data for $x \leq  0.051$ are published in Ref.
  \onlinecite{Kretzschmar2016}, while those with $x = 0.085$ are shown in Fig.~\ref{fig:BFCA85}.
  }
\label{fig:Lambda-chi}
\end{figure}

{Fig.~\ref{fig:Lambda-chi} shows $\beta\tilde\Lambda(\beta/2,T)$ and $\chi^\prime(0,T)$ extracted from the $B_{2g}$ Raman data of \Co122 using
Eqs.~(\ref{eq:kk},\ref{eq:betaover2}) for a range of doping concentrations,} $x$ below ($x\le0.051$) and above ($x=0.085$) a putative quantum critical point at $x_c\approx 0.06$. 
{ The scale on the left and right ordinates (for $\beta\tilde\Lambda(\beta/2,T)$ and $\chi^\prime(0,T)$ respectively) are chosen so that the two curves coincide at high $T$.}
For $x < 0.085$ the two quantities show a qualitatively similar temperature dependence above the structural transition temperature,
at which they both have a cusp singularity. However, as $x$ increases, the temperature dependence of $\beta\tilde\Lambda(\beta/2,T)$ weakens much more
rapidly than that of $\chi^\prime(0,T)$. 
The two measures show meaningfully distinct behavior at $x=0.085$, where $\chi^\prime(0,T)$ increases by nearly a factor of two upon cooling to 50\,K, while $\beta\tilde\Lambda(\beta/2,T)$ remains constant.

{ It is important to stress that 
there is an unavoidable uncertainty in the inferred values 
 of $\chi^\prime$.}  
Specifically, since
$\chi^{\prime\prime}(\omega,T)$ is essentially constant at high energies (see Fig.~\ref{fig:BFCA85}), to compute
$\chi^\prime(0,T)$ one must cut off the Kramers-Kronig integral, in which case the result depends logarithmically on the cutoff. {A corollary of this is that the degree of temperature dependence of $\chi^\prime$ depends strongly on the cutoff.} 
In contrast, the weighting factor $[\sinh(\beta\omega/2)]^{-1}$ in
Eq.~(\ref{eq:betaover2}) decays exponentially, making the integral unique so long as the spectra are measured up to energies of a few
times the temperature. In any case, as anticipated above, $\chi^\prime(0,T)$ and $\beta\tilde\Lambda(\beta/2,T)$ have
near-identical singularities at the structural transition temperature $T_s$ in underdoped \Co122 with $x\lesssim 0.06$. The
lack of a genuine divergence at the transition is likely an effect of electron-phonon coupling\cite{gallais2016}.

\section{Discussion}

In this paper, we discuss a method to analyze experimental spectroscopic data by transforming it to imaginary time.  
This method is applicable to almost any experimental probe which measures response functions at frequencies
of order the temperature (for additional examples see Appendix~\ref{sec:examples}). In particular, the appropriate
response function at maximal imaginary time separation, $\beta\tilde\Lambda(\beta/2)$, can be computed without an arbitrary
cut-off procedure, and is a quantitative measure of low frequency spectral weight.  For the optical conductivity,
$\beta\tilde\Lambda(\beta/2)$ is a physically motivated definition of a low frequency ``Drude weight''.\cite{lederer2017} In inelastic
neutron scattering, a drop in $\beta\tilde\Lambda(\beta/2)$ as a function of temperature can quantify the development of a spin gap. In angle resolved
photoemission spectroscopy,  
$\beta\tilde\Lambda(\beta/2)$ is a proxy for
the quasiparticle residue $Z$.~\cite{schattner2016} 
The
experimental measurement of imaginary time response functions is a potentially powerful tool both for the quantification of low
frequency spectral properties, and for bridging experiment and theory. 

{
Framing the analysis in terms of $\beta\tilde\Lambda(\beta/2,T)$  has three advantages: (i) 
{ This quantity} can be
computed directly and unambiguously from the measured $\chi^{\prime\prime}$; (ii) 
  it can be directly compared with theoretical predictions performed in the imaginary time domain
 \cite{schattner2016}; (iii) it 
 { highlights asymptotic low-energy} physics by suppressing the effects of high energy spectral features.}
 
{This last point is 
{vividly} illustrated by considering the Raman data  in Fig.~\ref{fig:Lambda-chi}d (\Co122 with $x=0.085$). A constant
value of $\beta\tilde\Lambda(\beta/2,T)$ indicates that $\chi^{\prime\prime}$ exhibits
$\omega/T$ scaling in a range of frequencies { 
that extends to well above $\omega=T$.
Such
scaling  up to a microscopic cutoff scale, $\Omega$, is  
a hallmark of  the  ``marginal Fermi liquid" phenomenology (see
Sec.~\ref{Asec:MFL}).  In this case, $\chi^\prime(0,T)$ would 
be expected to have a weak (logarithmic) $T$ dependence, 
while deviations of $\beta\tilde\Lambda(\beta/2,T)$ from a constant
value would be 
small for  $T\ll \Omega$.} 
Accordingly, the
temperature dependence of  $\beta\tilde\Lambda(\beta/2,T)$ suggests an
intermediate asymptotic range of singular behavior in $\chi^{\prime\prime}$ in a range of
frequencies and temperatures $T_c \ll \omega, T \ll \Omega$, while the temperature dependence
of $\chi^\prime(0,T)$ does not clearly manifest such behavior. 
{(The extent to which the indicated scaling is actually obeyed is exhibited in Fig. \ref{scaled}.)}
Thus, the imaginary time analysis is particularly suited to reveal the emergent $\omega/T$ scaling behavior at low frequencies.}

{
Indeed, there is abundant evidence\cite{Chu2012divergent,Gallais2013,Bohmer2014,kuo2016,Blumberg2016,Palmstrom2019} for nematic fluctuations near a putative quantum critical point (QCP) in  \Co122 at $x=x_c \approx 0.06$.  This would imply the existence of  a quantum critical fan in the $x-T$ plane that is bounded by crossover lines, $T^*(x) \sim |x-x_c|^y$ (where $y$ is an appropriate critical exponent).   While these considerations are only precise asymptotically close to the putative QCP, suggestive evidence of the existence of such a crossover scale is apparent in Fig. \ref{fig:Lambda-chi}.  In particular, $\beta\tilde\Lambda(\beta/2,T)$  for $x=0.085$ is approximately constant in the B$_{2g}$ channel (as indicated by the dotted line in the figure) until it deviates downward below $T^* \approx 60$K, while for $x=0.051$ it rises (as ``classical'' critical fluctuations associated with the approach to the ordered phase become significant) below $T^* \sim 150$K.  This, we feel, is a clear example of a way in which the present mode of analysis can lead to new ways to interpret data;  whether what  is at play is truly quantum critical nematic fluctuations can be tested by obtaining data closer to criticality, both by studying samples with $x$ closer to $x_c$ and, by suppressing superconductivity with a magnetic field, following the behavior to lower $T$.
}

\begin{acknowledgments}
 We acknowledge helpful discussions with A. Baum, M. Randeria, R. Scalettar, and N. Trivedi. D.J. and R.H. gratefully acknowledge the hospitality of the the Stanford Institute for Materials and Energy Sciences (SIMES) at Stanford University and SLAC National Accelerator Laboratory. Financial support for the work came, in part, from the Friedrich-Ebert-Stiftung (D.J.), the Deutsche Forschungsgemeinschaft (DFG) via the Priority Program SPP\,1458 (D.J., T.B., and R.H. project no. HA\,2071/7-2), the Collaborative Research Center TRR\,80 (D.J. and R.H., Project ID 107745057), and the Bavaria California Technology Center BaCaTeC (S.A.K., D.J., and R.H., project no. 21[2016-2]). S.A.K. was supported in part by NSF grant $\#$ DMR-1608055 at Stanford, S.L.  was supported by a Bethe/KIC fellowship at Cornell, and E.B. was supported by the European Research Council under grant HQMAT ($\#817799$) and by the Minerva foundation.

\end{acknowledgments}

\begin{appendix}
\section{Derivation of Equation \ref{basic}}
We begin with the expressions for $\tilde \chi_{ab}^{\prime\prime}(t)$ and $\tilde \Lambda_{ab}(\tau)$ in Lehmann representation:
\eq{
 \tilde \chi_{ab}^{\prime\prime}(t)=&\frac{1}{2Z}\sum_{n,m}\Phi_{a,nm}\Phi_{b,mn}\left[e^{-\beta E_n}-e^{-\beta E_m}\right] \\
\tilde \Lambda_{ab}(\tau)=&\frac{1}{Z}\sum_{n,m}\Phi_{a,nm}\Phi_{b,mn}e^{-\beta E_n}e^{\tau(E_n-E_m)} 
}

Fourier transform $\tilde\chi''(t)$ with appropriate regularization at $t\to\pm\infty$, yielding
\eq{
\chi^{''}_{ab}(\omega)=&\frac{1}{2Z}\sum_{n,m}\Phi_{a,nm}\Phi_{b,mn}\left[e^{-\beta E_n}-e^{-\beta E_m}\right]e^{it(E_n-E_m)}\nn\\
&\quad \times\frac{2\cdot0^+}{(\omega+E_n-E_m)^2+(0^+)^2}\nn\\
=&\frac{1}{2Z}\sum_{n,m}\Phi_{a,nm}\Phi_{b,mn}e^{-\beta E_n}\left[1-e^{-\beta\omega}\right]\nn\\
&\quad \times2\pi\delta(\omega+E_n-E_m),
}
with $0^+$ a positive infinitesimal. We can now write $\tilde\Lambda_{ab}(\tau)$ in terms of $\chi''_{ab}(\omega)$
\eq{
\tilde \Lambda_{ab}(\tau)=&\int \frac{d\omega}{\pi} \chi''_{ab}(\omega)\left[\frac{\exp(-\omega\tau)}{1-e^{-\beta\omega}} \right]\nn\\
=&\int \frac{d\omega}{2\pi} \chi''_{ab}(\omega)\left[\frac{\exp[-\omega(\tau-\beta/2)}{\sinh(\beta\omega/2)}\right],
}
recovering Eq.~\ref{basic}.
\section{Example Transforms}
\label{sec:examples}
\subsection{Nearly constant $\sigma$}

The optical conductivity $\sigma^\prime (\omega)$ is generically an analytic function of $\omega$, in which case there is a formal way to express $\tilde \Lambda$ as follows: Starting from Eq.~(\ref{basic}) for the current-current correlator,
\bea
\tilde\Lambda(\tau)&=& \int \frac{d\omega}{2\pi} \sigma'(\omega) \frac{\omega \exp[\omega (\tau - \beta/2)]}{\sinh(\beta \omega/2)} \nonumber \\
&=& \sigma'(\partial_\tau) \int \frac{d\omega}{2\pi} \frac{\omega \exp[\omega (\tau - \beta/2)]}{\sinh(\beta \omega/2)} \nonumber \\
&=& {\pi T^2}\ \sigma^\prime\left( \partial_\tau\right)\  \sec^2\left[\pi T(\tau-\beta/2)\right],
\label{eq:sigma}
\eea
where $\sigma'(\omega) = \chi''(\omega)/\omega$ is the real part of the optical conductivity, and $\sigma'(\partial_\tau)$ is obtained by expanding $\sigma'(\omega)$ in powers of $\omega$ and replacing $\omega\rightarrow \partial_\tau$. 
If $\sigma^\prime (\omega)$ varies slowly as a function of $\omega$ on the scale of $T$, then a low order Taylor expansion in $\omega$ is adequate. Then
\bea
&&\tilde{\Lambda}(\tau) = {\pi T^2}\ \sigma^\prime\left(0 \right)\ \sec^2\left[\pi T(\tau-\beta/2)\right]\times \Big\{1  \nonumber \\
&&+\alpha_2 \sec^2\left[\pi T(\tau-\beta/2)\right][4-2\cos\left[2\pi T(\tau-\beta/2)\right] \nonumber \\
&&+ \ldots\Big\}
\eea
where
\be
\alpha_2 =(\pi T)^2\left[ \frac {\partial_\omega^2 \sigma^\prime}{\sigma^\prime}\right |_{\omega=0}\sim \left(\frac{\pi T}{\tilde\gamma}\right)^2,
\ee
and we have defined $\tilde \gamma$ as a measure of the ``width" of the conductivity, and  the expansion is reasonable so long as $\tilde \gamma \gg \pi T$.
\subsection{Sharply peaked $\sigma$}
If $\sigma'(\omega)$ is negligible except for frequencies $|\omega|\ll T$, the corresponding imaginary time correlator is
nearly constant in $\tau$, with polynomial corrections given by moments of $\sigma'(\omega)$. This can be seen by Taylor expanding the integration kernel in the first line of Eq.~(\ref{eq:sigma}) for $|\omega \beta|,|\omega \tau|\ll 1$:
\bea
&\tilde \Lambda( \tau) =\frac{NT}{\pi}\left(1-\frac{\gamma^2}{24T^2}+\frac{\gamma^2}{2}[\tau-\beta/2]^2+\dots\right),
\eea
where the total optical weight is $N\equiv \int \sigma^\prime(\omega)d\omega$, and the squared width of the peak is $\gamma^2=N^{-1}\int \omega^2 \sigma^\prime(\omega)d\omega$.
\subsection{Marginal Fermi liquid}
\label{Asec:MFL}
The Raman response of many strongly correlated electron fluids can be well approximated (below a high frequency cut-off) by the ``marginal Fermi liquid'' form
\be
\chi_{\rm MFL}^{\prime\prime}(\omega) = A \tanh(\beta \omega/2).
\ee
This same form  arises as the local susceptibility of a two-channel Kondo impurity and in various other contexts.  Transforming this expression to imaginary time yields
\be
\tilde{\Lambda} (\tau ) = \frac {AT}{\cos[\pi T(\beta/2-\tau)]}
\label{LambdaMFL}
\ee
where the divergences as $\tau \to 0$ and $\tau\to \beta$ are cut off at short imaginary times of order the inverse cut-off.

\subsection{Quantum-Critical Power Law}
Near a QCP obeying $\omega/T$ scaling, one expects order parameter correlations to have a power law form 
for imaginary times $\tau$ long compared to microscopic time-scales $\tau_0$ but short compared to the thermal time, $\beta$.  To make the analysis simple, consider a pure power-law form
\be
\tilde \Lambda(\tau) \approx C\left[ \frac1 {|\tau|^{x}} +\frac 1 { |\beta-\tau|^{x}} \right].
\label{lambdaqcp}
\ee
The divergences at $\tau=0,\beta$ would be regularized at an appropriate UV cutoff scale. This can be done, e.g., by replacing $|\tau|^{-x}$ with $(\tau^2 + \tau_0^2)^{-x/2}$, where $1/\tau_0$ is a high energy cutoff, and similarly for $ |\beta-\tau|^{-x}$.

Working backwards, we see that for $x<2$, the corresponding expression in real-time is
\be
\chi^{\prime\prime}(\omega)= \frac{C}{ T^{1-x}}\ F(\beta\omega),
\label{scaling}
\ee
where $F$ is the scaling function
\be
F(u) =\frac {\pi u } {\Gamma(x)} \left |  \frac 1 u\right|^{2-x}\left[1-e^{-|u|}\right] \ ,
\label{scalingF}
\ee
with the gamma function $\Gamma(x)=\int_0^\infty \, y^{x-1} e^{-y} dy$.

Even precisely at a QCP, one expects pure power-law behavior of $\tilde{\Lambda}$ only for times $\tau_0 \ll \tau \ll \beta$. More generally, $\tilde{\Lambda}$ near a QCP reads   
\be
\tilde{\Lambda}(\tau) \approx T^x f(\tau/\beta),
\label{scaling_lambda}
\ee
where $f$ is a scaling function, and the scaling form holds as long as $\tau\gg \tau_0$ and $\beta-\tau\gg \tau_0$. 
For example, the marginal Fermi liquid form in Eq. \ref{LambdaMFL} shows the same power-law behavior for $\tau\ll \beta/2$ as does Eq. \ref{lambdaqcp} with $x=1$, but differs from this expression for $\tau$ near $\beta/2$.
%

If $\tilde \Lambda(\tau)$ obeys Eq.~\ref{scaling_lambda} in the regime $\tau_0 \ll \tau \ll |\beta-\tau_0|$ then $\chi''(\omega)$ has the same scaling form as in Eq.~\ref{scaling}, but the scaling function $F$ depends on the behavior of $\tilde \Lambda(\tau)$ when $\tau \sim \beta/2$.   While the above expressions are pleasingly explicit, in the more general case,
if $\omega_1$ is a low frequency scale that measures the distance to the QCP (at which $\omega_1$ would vanish), 
  then the essential aspects of this analysis can be restated as
\be
\chi^{\prime\prime}(\omega) \sim \left[\frac {\omega} {|\omega|^{2-x}}\right]
\times \left\{
\begin{array}{ccc}
\beta |\omega|  & {\rm for}   &  \omega_1 \ll |\omega|\ll T \\
1  & {\rm for}   & T \ll |\omega| \ll 
\tau_0^{-1}
\end{array}
\right .
\ee
In particular, the critical exponent, $x$, governing the behavior of $\tilde{\Lambda}(\tau)$ for $\tau_0 \ll \tau \ll \beta$  determines  the frequency dependence of $\chi^{\prime\prime}$ both in the range $T \ll \omega \ll \tau_0^{-1}$, and in the range $\omega_1 \ll \omega \ll T$, but does not by itself give the relative value of the amplitudes.
\end{appendix}

\bibliography{final}

\begin{thebibliography}{16}
\bibitem [{foo()}]{footnote}%
  \BibitemOpen
  \href@noop {} {}\bibinfo {note} {We use the crystallographic two-iron unit
  cell. In the one-iron unit cell this would be B$_{1g}$.}\BibitemShut {Stop}%
\end{thebibliography}

\end{document}